\documentclass[
 reprint,
 amsmath,amssymb,
 aps,
]{revtex4-2}

\usepackage{graphicx}
\usepackage{subcaption} 
\usepackage{dcolumn}
\usepackage{bm}
\usepackage{hyperref}
\usepackage{xcolor}
\usepackage{url}

\begin{document}

\preprint{APS/123-QED}

\title{Contest Dynamics Between Cooperation and Exploitation}

\author{Alfonso de Miguel-Arribas}
\altaffiliation{These authors contributed equally to this work. The order of authorship was determined alphabetically by last name.}
\affiliation{
 Institute for Biocomputation and Physics of Complex Systems (BIFI), University of Zaragoza, 50018, Zaragoza, Spain
}
\affiliation{
 Zaragoza Logistics Center (ZLC), Av. de Ranillas 5, 50018, Zaragoza, Spain
}

\author{Chengbin Sun}
\altaffiliation{These authors contributed equally to this work. The order of authorship was determined alphabetically by last name.}
\affiliation{
  School of Economics and Management, Dalian University of Technology, Dalian 116024, China
}
\affiliation{
  Institute for Biocomputation and Physics of Complex Systems (BIFI), University of Zaragoza, 50018, Zaragoza, Spain
}

\author{Carlos Gracia-Lázaro}
\email{cgracial@usj.es}
\affiliation{Universidad San Jorge (USJ), Autovía Mudéjar, km. 299, 50830, Villanueva de Gállego, Zaragoza, Spain.}


\author{Yamir Moreno}
\affiliation{
 Institute for Biocomputation and Physics of Complex Systems (BIFI), University of Zaragoza, 50018, Zaragoza, Spain
}
\affiliation{
 Department of Theoretical Physics, University of Zaragoza, 50018, Zaragoza, Spain
}

\date{\today}

\begin{abstract}
Cooperation and competition are fundamental forces shaping both natural and human systems, yet their interplay remains poorly understood.  The Prisoner’s Dilemma Game (PDG) has long served as a foundational framework in Game Theory for studying cooperation and defection, yet it overlooks explicit competitive interactions. Contest Theory, in turn, provides tools to model competitive dynamics, where success depends on the investment of resources. In this work, we bridge these perspectives by extending the PDG to include a third strategy, \textit{fighting}, governed by the Tullock contest success function, where success depends on relative resource investments. This model, implemented on a square lattice, examines the dynamics of cooperation, defection, and competition under resource accumulation and depletion scenarios. Our results reveal a rich phase diagram in which cooperative and competitive strategies coexist under certain critical resource investments, expanding the parameter space for cooperation beyond classical limits. Fighters delay the dominance of defectors by mediating interactions, expanding the conditions under which cooperation persists. This work offers new insights into the evolution of social behaviors in structured populations, bridging cooperation and competition dynamics.
\end{abstract}


\maketitle

\section{Introduction}
\label{sec:intro}

Competition within, between, and among species is one of the most significant factors influencing the distribution and abundance of living organisms. In the real world, living beings, including humans, engage in competition for finite resources to continue their struggle for survival. Along with predation and mutualism, competition is one of the three major biological forces that assemble living communities \cite{keddy2012competition}. For humans, competition extends beyond basic survival needs, manifesting across diverse scales and domains, from personal relationships and workplace dynamics \cite{drago1991competition, jones2017competition, swab2019steel} to global economic markets \cite{porter1986competition, porter2000location, barnett2004some} and geopolitical conflicts \cite{markowitz2018power, flint2019geopolitics}. Examples include striving for career advancements \cite{freeman2001competition, johnsen2020competition}, competing for limited resources such as housing \cite{turnbull2006spatial, oxley2008competition, dunn2021rural} or education \cite{epple2002ability, hart2024competition}, and nations vying for technological or military supremacy \cite{liff2014racing, raska2019strategic}, and sports \cite{arnold1983three, mcgarry2002sport, furley2019modern}. On the other hand, cooperation (sometimes subsumed under the mutualism umbrella) is another significant behavior observed in life. Biologists are particularly interested in cooperation because it appears to contradict the competitive nature fundamental to natural selection. Why should one individual help another, especially when they are competitors in the struggle for survival? Yet, cooperation is abundant in nature \cite{axelrod1981evolution, sachs2004evolution, boyd2009culture}. 

The Prisoner's Dilemma (PD) stands out as a foundational model in both classical and evolutionary game theory. Together with other social dilemmas, it has been extensively studied to understand the mechanisms under which cooperation can naturally emerge and persist \cite{rand2013human, perc2017statistical}. This game encapsulates a conflict between what is optimal from an individual’s perspective and what benefits the collective, illustrating the tension between cooperation and exploitation between self-interested individuals \cite{axelrod1981evolution}. This conflict jeopardizes nearly every form of cooperation, including trade and mutual aid \cite{nowak2000cooperation}. This simple yet profound thought experiment has generated extensive research in diverse disciplines, including economics \cite{von2007theory, conybeare1984public, rubinstein1986finite, clark2001sequential}, political science \cite{snyder1971prisoner, axelrod1980effective, snidal1985coordination}, biology and ecology \cite{smith1973logic, rapoport1989prisoner, dugatkin1992beyond, doebeli2005models}, and statistical physics \cite{szabo1998evolutionary, szabo2007evolutionary, perc2017statistical}. 

Contest Theory, on the other hand, is another branch of Game Theory and economics that focuses on competition and competitive strategies \cite{vojnovic2015contest, corchon2018contest}. Its origins trace back to seminal contributions by Tullock \cite{tullock1967welfare, tullock1980efficient} and Krueger \cite{krueger2008political}, who studied rent-seeking, and Becker \cite{becker1983theory}, who investigated lobbying. Since then, Contest Theory has been applied to a variety of domains, including elimination tournaments \cite{rosen1985prizes}, conflicts \cite{hirshleifer1989conflict,skaperdas1992cooperation}, political campaigns \cite{skaperdas1995modeling}, economic contests \cite{fahy2002role, szymanski2003economic}, sports \cite{nitzan1994modelling}, network attacks \cite{goyal2012competitive, goyal2014attack}, and wars \cite{acemoglu2012dynamic,baliga2013bargaining, caselli2015geography,krainin2016war,novta2016ethnic, dziubinski2021strategy}. 

Formally, a contest is characterized by a set of agents, their respective efforts, a tentative payoff for each contestant (the prize), and a set of functions, known as contest success functions (CSFs), that determine the individual probabilities of obtaining the prize based on the agents’ efforts \cite{skaperdas1996contest, corchon2018contest}.

The versatility of the Prisoner's Dilemma Game (PDG) has led to numerous variations and extensions, each aimed at capturing different aspects of strategic behavior in complex systems. A particularly fruitful approach has been the study of the PDG, as well as other games, through the lens of evolutionary game theory on structured populations, such as lattices and complex networks \cite{szabo2007evolutionary}, using methods from statistical physics \cite{perc2017statistical}. Extensions of the game have incorporated additional strategies beyond the elementary cooperation $C$ and defection $D$, such as tit-for-tat ($T$) \cite{nowak1993strategy, szabo2000spatial}, which involves responding to others' actions reciprocally (cooperation or defection, depending on context). Other notable strategies include the loner strategy ($L$), which allows for voluntary withdrawal from interactions \cite{szabo2002evolutionary, jia2018impact}, and the exit strategy ($E$), where players leave the game with a small fixed payoff \cite{orbell1984cooperators, phelan2005using, seale2006modeling, shen2021exit}. While defection may be regarded as a competitive stance against peers, it is often seen as exploitative or unfair. By contrast, competition as posed by Contest Theory emphasizes a “fair play” perspective, where success is determined by invested resources relative to others and any player has a finite chance of winning.

To the best of our knowledge, no extension of the PDG or other social dilemmas has introduced a purely competitive strategic behavior as conceived by Contest Theory. Given the ubiquity of competition in nature and human societies, this work proposes an evolutionary game that combines the standard PDG with explicit resource-based competition as conceptualized in Contest Theory. Specifically, we introduce a third strategy, referred to as ``fighting'' ($F$), and model interactions involving this strategy using the Tullock CSF, which quantifies the probability of success based on the invested resources in the contest \cite{tullock1980efficient, tullock2008efficient}. This extension, which we refer to as the cooperation-defection-fighting (CDF) model, allows for a more comprehensive exploration of conflict and competition dynamics, incorporating fundamental strategic behaviors observed in both natural and social systems. For simplicity, the game is played on a standard regular square lattice, considering scenarios with infinite resource accumulation and depletion.

Our results reveal a phenomenologically rich phase diagram, highlighting the coexistence of cooperative and competitive strategies when a critical resource investment is made in contests. Interestingly, beyond a certain investment, the surviving fighter population diminishes again, unfolding another nonlinearity in the system's behavior, now related to the optimal population of competitors. Furthermore, the co-existence region between cooperation and competition extends beyond the classical critical value of the temptation parameter $b$, causing the dominance of defectors to be delayed to higher values.

This work is organized as follows. Section \ref{sec:model} introduces the proposed extension of the PDG incorporating the competitive \textit{fighter} strategy. Section \ref{sec:results} presents the results, describing the system's macroscopic behavior, the nature of the phase transition, and the optimality of the fighter population. Finally, Section \ref{sec:discussion} concludes with a discussion of the implications.

\section{CDF Model}
\label{sec:model}

We introduce the CDF model as an evolutionary $3$-strategy game with players located on the vertices of a $2$-dimensional regular square lattice of size $L\times L$ with periodic boundary conditions. The total number of players is $N=L\times L$. Each player, or agent, is allowed to interact only with their four nearest neighbors in a pairwise fashion, and self-interactions are excluded.

\subsection{Strategies and payoff matrix}
\label{subsec:strategies}

 The following strategies are considered:
\begin{itemize}
    \item Cooperator ($C$). Individuals choosing to cooperate engage in mutually beneficial interactions.
    \item Defector ($D$). Individuals choosing to defect prioritize their payoff at the expense of their interacting neighbor.
    \item Fighter ($F$). Individuals selecting to fight engage in direct contests with their neighbors, aiming to obtain resources through competition.
\end{itemize}

The payoff matrix informs us about the outcome of every potential interaction in the system and for the proposed $3$-strategy game is given by:
\begin{equation}
    \begin{array}{c|ccc}
        & C & D & F \\
        \hline
        C & 1 & 0 & |G(r_i, r_j)| \\
        D & b & 0 & |G(r_i, r_j)| \\
        F & |G(r_i, r_j)| & |G(r_i, r_j)| & |G(r_i, r_j)|
    \end{array}
\end{equation}
Here, $b>1$ is commonly referred to as the temptation to defect or, simply, the temptation parameter, and thus the subset of interactions involving cooperators and defectors constitutes the weak Prisoner's Dilemma (PD) \cite{rapoport1988experiments}. The parameter $b$ will be treated as one of the control parameters in our analysis. On the other hand, the function $G(r_i,r_j)$ represents the result of the contest that takes place between agents $i$ and $j$ in interactions in which at least one of the contenders involved is a fighter. The quantities $r_{i}$ and $r_{j}$ are the resources of players $i$ and $j$, respectively, invested in the contest. The specific sign of $G(r_i,r_j)$ ($\pm$) will depend on the outcome resolution. Next, we offer a complete description of the competitive contest.

\subsection{Contest resolution}
\label{subsec:contest}

Players invest a certain fraction of their resources when facing a contest or fighting interaction. For an agent $i$, the invested resources are defined as:
\begin{equation}
    r_i\equiv\rho\frac{\Pi_i}{n_F},
\end{equation}
where $\Pi_i$ are the total resources of agent $i$, $n_F$ is the number of contests agent $i$ will have to face in the corresponding time step or round, and $\rho$ is a control parameter that represents the fraction of their total resources that agents are willing to invest. We will refer to $\rho$ as the investment fraction. This is one of the control parameters in our model and will be assumed identical for all the players in the system. Note that $\Pi_i$ and $n_F$ are stochastic dynamical variables subjected, respectively, to resource updating resulting from the game interactions, and to the strategy updating of agents.

Now, how are contests resolved and, subsequently, invested resources redistributed? We use the standard Tullock contest success function (CSF) from Contest Theory \cite{tullock1980efficient, tullock2008efficient}, which reads: 
\begin{equation}
    P_{CSF}(r_i,r_j;\gamma)=\frac{r_i^\gamma}{r_i^\gamma+r_j^\gamma}.
\end{equation}
The Tullock function yields the probability of agent $i$ beating agent $j$ in a contest, given they invest, respectively, $r_i$ and $r_j$ resources. The larger the invested resources, the higher the winning probability. The Tullock CSF has a free parameter $\gamma$, $0\leq\gamma$, which modulates the influence of the resources invested in the winning probability and has been referred to as the technology parameter \cite{dziubinski2021strategy} or elasticity \cite{he2024complexity}. When $\gamma\geq 1$, we speak of a rich-rewarding regime, whereas when $\gamma<1$, we enter the poor-rewarding regime \footnote{Note that the winning probability of any fighter is invariant under the re-scaling of both contenders’ resources, i.e., provided $\rho>0$, the probability $P_{CSF}(i\to j;\gamma)$ is independent of the fighting investment factor $\rho$; nevertheless, it has to be highlighted that the resource distribution after the combat strongly depends on $\rho$.} (see Appendix \ref{app:tullock} for more details on the Tullock CSF behavior).

Without loss of generality, assuming agent $i$ wins a contest, the winner takes all the invested resources and therefore:
\begin{equation}
G(r_i,r_j)=
\begin{cases}
    +r_j & \text{for agent $i$,} \\
    -r_j & \text{for agent $j$.}
\end{cases}
\end{equation}
Contests then behave as zero-sum games that redistribute the wealth of the players.

\subsection{Dynamics}
\label{subsec:dynamics}

We always assume an equal initial fraction of every strategy in the evolutionary game. Let $f_X=N_X/N$ be the fraction of individuals with strategy $X=\{C,D,F\}$, then: 
\begin{equation}
    f_C(0)=f_D(0)=f_F(0)=\frac{1}{3}.
\end{equation}
Every player in the game is endowed with an initial resource quantity equal to unity:
\begin{equation}
    \Pi_i(t=0)=1\;\;\;\forall i\in N. 
\end{equation}
Payoffs/resources obtained through the different interactions are accumulated round after round, and thus we may refer to $\Pi_i(t)$ as the player's $i$ cumulative resources or cumulative payoff at time $t$. In contrast, payoffs accumulated in a single, independent round are named round or instant payoffs, denoted by $\pi_i(t)$.

For the simulation of this game's dynamics, a discrete-time evolution with synchronous updating is assumed. At every round (time step), every individual interacts with all four nearest neighbors. We make sure that every pairwise interaction occurs only once per round. Within every round $t$, each agent accumulates a certain round or instant payoff $\pi_i(t)$, whose value will depend on the outcome of the specific interactions of the focal agent with their neighbors. The upper bound of this quantity is $+4$ if all the neighbors are cooperators and so is the focal agent, whereas the lower bound is $-r_i$ in the event that the focal agent is a fighter and/or all their neighbors are fighters, and the focal agent loses every contest. Thus, depending on the focal agent's strategy and their neighborhood, a varied landscape for the instant payoff is expected. In any case, the resulting $\pi_i(t)$ accumulates to the agent's cumulative resources as $\Pi_i(t)+\pi_i(t),\;\forall i\in N$.

After all interactions have taken place for every player in the system, they proceed to review their strategies for the next round of the game. As it is customary in evolutionary game theory, we use the Fermi updating rule. Each agent $i$ chooses a random neighbor $j$. Agent $i$ imitates the current round strategy (time $t$) of agent $j$, $\sigma_i\leftarrow \sigma_j$, with a probability given by
\begin{equation}
    P_F(\sigma_i\leftarrow \sigma_j)=\frac{1}{1+\exp[(\Pi_i-\Pi_j)/\kappa]},
\end{equation}
where $\sigma_{i(j)}$ and $\Pi_{i(j)}$ represent, respectively, the current strategy and cumulative payoff of agent $i$ ($j$), and $\kappa$ is the noise parameter modulating the randomness or rationality of the decision-making process. Here, $\kappa=0.1$ is taken as a fixed parameter. 

We introduce a further novelty in our model by adding a depletion term to the cumulative resources of players by the end of each round. Beyond helping control explosive resource growth in the model, this mechanism also represents an inescapable thermodynamic feature of nature, affecting both individuals and more complex systems. Thus, to account for eventual resource depletion, each player's resources evolve over time as:
\begin{equation}
    \Pi_i(t+1)=(1-\alpha)[\Pi_i(t)+\pi_i(t)]\;\;\;\forall i\in N,
\end{equation}
where $\pi_i(t)$ depends on each particular agent's interactions at each round. The parameter $\alpha$ is the depletion rate and is assumed to be equal for every individual in the system. Clearly, if $\alpha=0$, resources accumulate indefinitely, whereas if $\alpha=1$, agents lose all their resources from one round to the next one, akin to a situation without any sense of memory. Note that for $\rho=0$ and $\alpha=1$, the game is equivalent to the voluntary Prisoner’s Dilemma, since there is a fraction of players, $f_F$, abstaining from participating. Additionally, we recover the weak PD for $\rho=0$, $\alpha=1$, and an initial fraction of fighters equal to zero, $f_F(0)=0$. We fix $\alpha=0.1$ throughout our whole analysis, and we will show some other scenarios for $\alpha$ in the supplementary material (See Appendix \ref{app:depletion}).

Finally, we characterize the state of the system macroscopically as given by the vector:
\begin{equation}
    \textbf{f}(t)=\left[f_C(t),f_D(t),f_F(t)\right],\;\;\;t\geq 0,
\end{equation}
informing us about the fraction of adopters of any strategy at every round $t$ of the evolutionary dynamics. Depending on the region in control parameter space generated by $(b,\rho,\gamma,\alpha)$, one or more strategies may survive in the long run, that is, at $t\to\infty$. If only a strategy exists after a transient time, effectively the dynamics reach an absorbing state, whereas otherwise, strategies coexist around some equilibrium point. 

\subsection{Payoffs: analytical considerations}
\label{subsec:payoffs}

Cooperators are the primary producers of wealth, while defectors, when exploiting cooperators, also contribute to wealth production, albeit to a lesser extent, as it ultimately originates from cooperators. Fighters, on the other hand, solely redistribute wealth.

The upper bound for the cumulative resources corresponds to a scenario where all agents cooperate, that is, $f_C(0)=1$, and therefore, $f_C(t)=1$ for any $t>0$. In that case, we have this relation for the cumulative payoff $\Pi_{C,n_C=4}$ of a cooperator surrendered by $n_C=4$ cooperators (and no fighters, nor defectors):
\begin{equation}
\Pi_{C,4}(t+1)=(1-\alpha)\left[\Pi_{C,4}(t)+k\right],
\end{equation}
where $k$ is the agent's connectivity, which for our structure setup is $k=4$ for everyone. Note that this is just the round payoff $\pi_i(t)=k$ in this situation. Now, taking the limit $\Delta t\to 0$ to continuous time, $\Pi_{C,4}$ evolves according to the differential equation:

\begin{equation}
\frac{d\Pi_{C,4}(t)}{dt}=-\alpha \Pi_{C,4}(t)-\alpha k +k,
\label{eq_fin_diff_payoff_CC}
\end{equation}
whose solution is
\begin{equation}
\Pi_{C,4}(t) = \frac{k}{\alpha}-K+e^{-\alpha t}.
\end{equation}

Therefore, the stationary limit for the upper bound of the cumulative payoff for the entire system, $\Pi_{f_C=1}(t)$, is:
\begin{equation}
\lim_{t \rightarrow \infty} \Pi_{f_C=1}(t) = N \lim_{t \rightarrow \infty} \Pi_{C,4}(t) = Nk\frac{1-\alpha}{\alpha}\; .
\end{equation}
As described before, the case of $\alpha=1$ implies no memory of the past, i.e., no accumulation, whereas the case of $\alpha=0$ drives unbounded resource growth. 

Clearly, the introduction of defectors, even yielding a $D$-$C$ interaction payoff of $b>1$, requires indeed the presence of cooperators, and thus these exploited cooperators underperform in their harvested round payoff $\pi<4$. Consequently, far from the obtained upper bound of a full cooperation configuration.

At the other extreme, the lower bound for the cumulative payoff corresponds to a scenario without cooperators, $\Pi_{f_C=0}$. Indeed, in the absence of cooperators, defectors cannot exploit them and no source of wealth exists. Thus, fighters and defectors alike compete for an ever-diminishing pool of resources. Independent of whether some players fare better than others,
\begin{equation}
    \lim_{t\to\infty}\Pi_{f_C=0}(t)\to 0,\;\;\;\text{provided}\;\;\alpha\neq 0.
\end{equation}
If, on the other hand, resource depletion does not apply, $\alpha=0$, and no cooperators are present in the system, the global pool of initially assigned resources ($\sum_{i=1}^N1$) is conserved, and thus $\Pi_{f_C=0}(t\to\infty)=N$. Under a homogeneous system in terms of connectivity and initial resource allocation, resource inequality among players will be just the result of pure chance and, over all, the resource-rewarding contest function.

\subsection{Simulations}
\label{subsec:simulations}

The dynamics described previously will be simulated through a microscopic stochastic simulation or multi-agent system simulation, as it is sometimes referred to in the evolutionary games literature, under a discrete-time and synchronous updating scheme. 

The following parameters will remain fixed and equal to the following values unless otherwise specified. Lattice dimension $L=100$, number of agents $N=L\times L=10^4$, noise parameter $\kappa=0.1$, and depletion rate $\alpha=0.1$. The main control parameters will be the temptation to defect $b$, with range in $b\in[1,2]$, and the invested fraction of resources in contest $\rho$, with $\rho\in[0,1)$. We will also explore some scenarios for varying technology parameter $\gamma$, focusing on the region $[0,1)$ (See Appendix \ref{app:technology} for the case $\gamma>1$).

Given the stochastic nature of the simulations, a transient or thermalization time $t_{\text{th}}=O(10^4)$ is given for the evolutionary dynamics to reach an equilibrium state and then, relevant observables are averaged over a certain time window, $t_{\text{obs}}=O(10^3)$, when more than one strategy coexists. Moreover, these mean values per simulation are further averaged over an ensemble of simulations run under the same set of control parameters, with $N_{\text{sims}}=10^2$.

The code developed for the simulations, analysis of results, and figure generation is hosted at \cite{phononautomata2025coopfight}. 

\section{Results}
\label{sec:results}

As previously indicated, we characterize the system's macroscopic behavior by the fraction of players under each possible strategy at equilibrium as given by the state vector $\textbf{f}(t)=[f_C(t),f_D(t),f_F(t)]$ at $t\to\infty$. The control parameter space to explore the system's behavior is exhaustively scanned in the region $(b,\rho)$ for specific values of $\gamma$, and $\alpha=0.1$ is always fixed unless otherwise specified. The elements of the state vector are averaged over a time window $t_{\text{obs}}$ for a given realization and over an ensemble of stochastic realizations $N_{\text{sims}}$. These values therefore, are computed as
\begin{equation}
    \langle f_X(\infty;b,\rho)\rangle=\frac{1}{N_{\text{sims}}}\frac{1}{t_{\text{obs}}}\sum_{s=1}^{N_{\text{sims}}}\sum_{t>t_{\text{th}}}f_{X;s}(t;b,\rho),
\end{equation}
with $f_{X;s}(t)$ the population fraction of strategy $X=\{C,D,F\}$, at round $t$ for the stochastic realization $s$. For brevity, we will refer to these quantities simply as the fraction of the given strategy, even though they are clearly averaged values. We also drop the reference to parameters $(b,\rho)$ when referring in general to these strategy population fractions.

\subsection{Macroscopic behavior}
\label{subsec:macroscopic}

\begin{figure*}
\centering
    \includegraphics[width=1.0\textwidth]{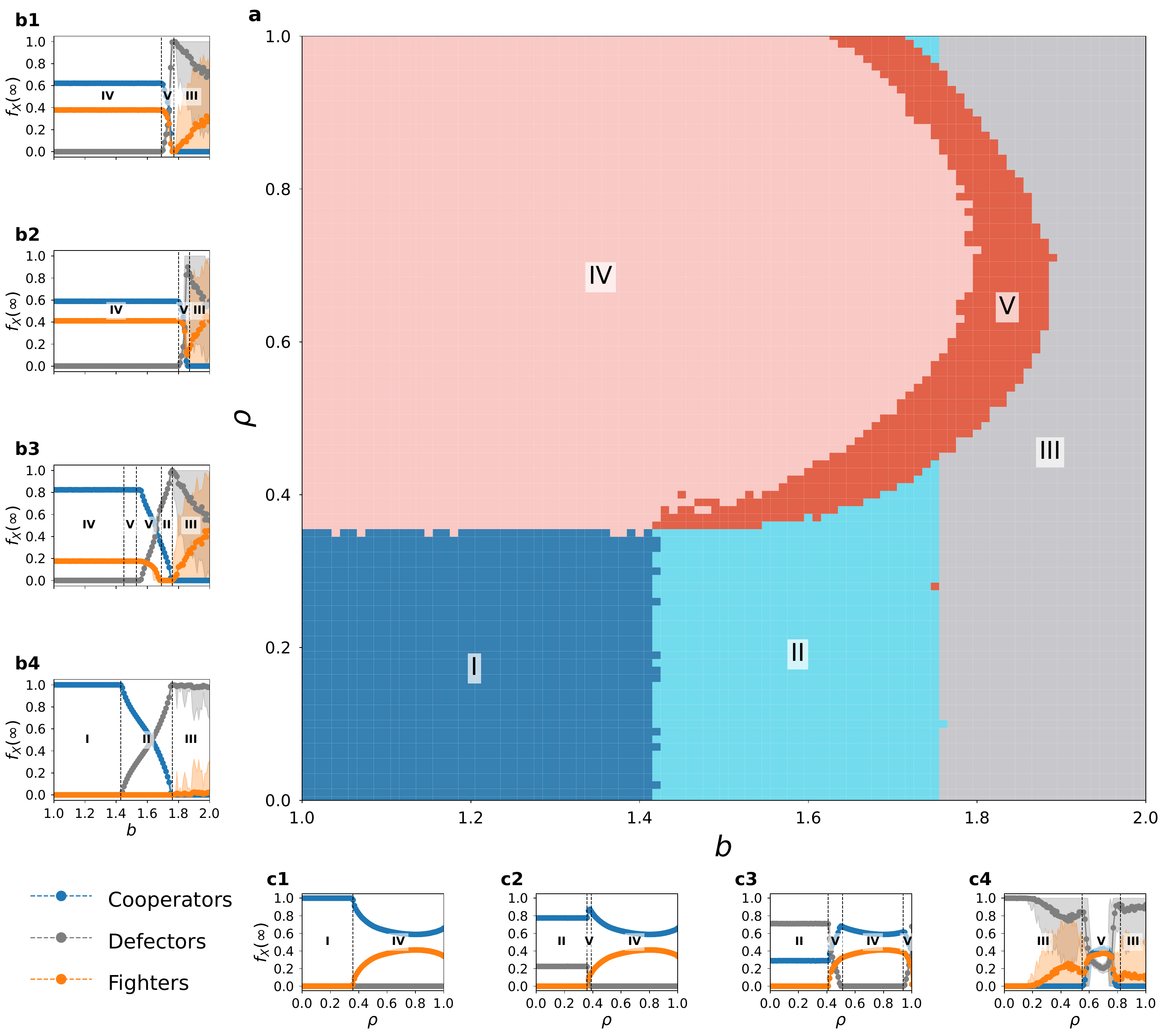}
    \caption{\textbf{Representative macroscopic behavior at equilibrium.} The central panel (a) depicts the system's macroscopic phases in the temptation-to-defect and investment fraction $(b,\rho)$ parameter space, for fixed technology parameter $\gamma=0.45$ and cumulative resources depletion rate $\alpha=0.1$. Phases are defined based on the macroscopic state vector at equilibrium, $\textbf{f}(\infty)=\left[\langle f_C(\infty)\rangle,\langle f_D(\infty)\rangle,\langle f_F(\infty)\rangle\right]$, where elements correspond to the average equilibrium fractions of cooperators, defectors, and fighters, respectively. Depending on the vanishing of some or none of these fractions, distinct phases emerge. Phase I: Full Cooperation, $\langle f_D(\infty)\rangle=\langle f_F(\infty)\rangle=0$. Phase II: Cooperation and defection coexistence. Phase III: Dead economy, characterized by $\langle f_C(\infty)\rangle=0$. Phase IV: Cooperation and fighting coexistence, with defectors absent. Phase V: Triple coexistence of cooperators, defectors, and fighters. Each point in the phase diagram results from averaging over $100$ simulations with system size $N=10^4$. The left column panels (b1–b4) display equilibrium strategy fractions as a function of $b$ for fixed $\rho$ values: $\rho=0.9$ (b1), $0.7$ (b2), $0.4$ (b3), and $0.2$ (b4). The bottom row panels (c1–c4) present fractions as functions of $\rho$ for fixed $b$ values: $b=1.1$ (c1), $1.5$ (c2), $1.7$ (c3), and $1.9$ (c4). Dashed lines indicate phase transition boundaries. These results are consistent with the persistence of fighters in Phase II, as suggested by local configuration dynamics discussed in the main text.} \label{fig:phases_master}
\end{figure*}

Through extensive Monte Carlo simulations of the proposed three-strategy evolutionary game, we uncover a rich phase structure. Figure \ref{fig:phases_master} illustrates a representative case of this macroscopic diversity, revealing up to five distinct equilibrium phases, characterized by the average macroscopic state vector defined as $\textbf{f}(\infty)=\left[\langle f_C(t)\rangle,\langle f_D(t)\rangle,\langle f_F(t)\rangle\right]|_{t\to\infty}$. The delimitation of phases is based on whether one, several, or none of the strategy fractions vanish at equilibrium. Accordingly, we define the phases discovered as follows:
\begin{itemize}
    \item \textbf{Phase I: Full Cooperation.} Cooperation becomes not only dominant but exclusive, with $\langle f_C(\infty)\rangle=1$ and $\langle f_D(\infty)\rangle=\langle f_F(\infty)\rangle=0$. This phase occurs for $b\in[1,b_c)$, following classical Prisoner's Dilemma behavior, and for $\rho<\rho_c$, where investment in fighting is insufficient to sustain fighters in the long run.

    \item \textbf{Phase II: Cooperation and Defection.} When $b$ exceeds $b_c$, defection eventually begins to erode cooperation. This phase, characterized by the coexistence of cooperators and defectors with $\langle f_F(\infty)\rangle=0$, mirrors standard PDG dynamics. For high enough $b$, defection overtakes cooperation and approaches hegemony.

    \item \textbf{Phase III: Dead Economy (No Cooperation).} For sufficiently high $b$, cooperation collapses entirely, $\langle f_C(\infty)\rangle=0$. Although defectors and fighters may persist, the absence of cooperators, the sole agents generating wealth through $C$-$C$ interactions, drives the system toward resource exhaustion. Consequently, the individual payoffs $\Pi_i(t\to\infty)\to 0$ for all players ($\sigma_i=D$ or $F$). Interestingly, this phase does not emerge from $\rho=0$ but from a somewhat higher value.

    \item \textbf{Phase IV: Cooperation and Competition.} At sufficiently large $\rho>\rho_c$, fighters (true competitors) survive and coexist with cooperators at equilibrium, while defectors (exploiters) are eliminated, $\langle f_D(\infty)\rangle=0$. Remarkably, this phase persists even for $b>b_c$, extending up to $b\approx 1.8$ for $\rho\approx 0.7$, and bordered by Phase V, which partially envelops Phase IV through a curved, semi-circular boundary.

    \item \textbf{Phase V: Triple Coexistence.} In a narrow, semi-circular region interfacing Phases III and IV, all three strategies are able to coexist. Here, cooperators tend to constitute the majority among the population, while fighters and defectors alternate as the second most abundant strategy, often running close to each other in fraction. This phase occurs at intermediate-to-high $b$ and large $\rho\in(\rho_c,1]$.
\end{itemize}

In addition to the macroscopic phase diagram, Figure \ref{fig:phases_master} presents detailed cross-sections of the equilibrium strategy fractions to elucidate transitions across control parameters.

Specifically, the bottom-row panels (c1–c4) show the equilibrium fractions of cooperators, defectors, and fighters as functions of $\rho$ for fixed $b$ values: $b=1.1$ (c1), $1.5$ (c2), $1.7$ (c3), and $1.9$ (c4).

Starting with panel c1, as $\rho$ increases, the system transitions from Phase I to Phase IV: a finite fraction of fighters survives and coexists with cooperators beyond the critical threshold $\rho_c\approx0.4$. The fighter population exhibits a non-monotonic dependence on $\rho$, reaching a maximum at $\rho\approx 0.8$ with $\langle f_F(\infty)\rangle>0.4$, signaling the existence of an optimal investment level $\rho_{\text{opt}}$. Beyond $\rho_{\text{opt}}$, higher investment in fighting diminishes the fighter fraction, rendering further investment detrimental. In panel c2 ($b=1.5$), for $\rho<\rho_c$, cooperation coexists with defection (Phase II). Increasing $\rho$ beyond $\rho_c$ leads again to cooperation-fighting coexistence (Phase IV). Here, $\rho_c$ remains approximately unchanged relative to panel c1, indicating that moderate increases in $b$ do not significantly shift the critical investment threshold. panel c3 ($b=1.7$) reveals a richer behavior. For $\rho<\rho_c$, defection dominates over cooperation due to the high temptation-to-defect payoff. As $\rho$ increases, a narrow window of triple coexistence (Phase V) emerges before returning to Phase IV. In Phase V, defectors survive as a minority alongside cooperators and fighters, but their presence remains marginal. Further increasing $\rho$ strengthens fighters while eliminating defectors, until eventually fighters themselves collapse at large $\rho$, returning the system to defector dominance. Finally, panel c4 ($b=1.9$) shows a Phase II where cooperation is close to extinction, and Phase III enters and dominates throughout this cross-section, interrupted by Phase V. 

The left-column panels (b1–b4) depict the complementary behavior, representing equilibrium strategy fractions as functions of $b$ for fixed $\rho$ values: $\rho=0.9$ (b1), $0.7$ (b2), $0.4$ (b3), and $0.2$ (b4). Starting with panel b4 ($\rho=0.2$), the system remains in full cooperation (Phase I) until $b$ exceeds a critical value $b_c$. Beyond this point, cooperation collapses rapidly, and defection dominates, leading to Phase III (dead economy). In Panel b3 ($\rho=0.4$), the system initially resides in Phase IV, with cooperation and fighting coexisting. As $b$ increases, a region of triple coexistence (Phase V) emerges before defectors become dominant, resulting again in Phase III. In this case, defection and fighting are more balanced compared to the low-$\rho$ regime, although both strategies ultimately compete for a vanishing pool of resources. Panels b2 ($\rho=0.7$) and b1 ($\rho=0.9$) display qualitatively similar behaviors, but the boundaries of Phase V shift toward higher $b$ values as $\rho$ increases. Higher investment in fighting stabilizes cooperation against defection up to larger temptations, delaying the transition to the dead economy.

\subsection{Nonlinear phenomena involving competition}
\label{subsec:nonlinear}

\begin{figure*}
\centering
    \includegraphics[width=1.0\textwidth]{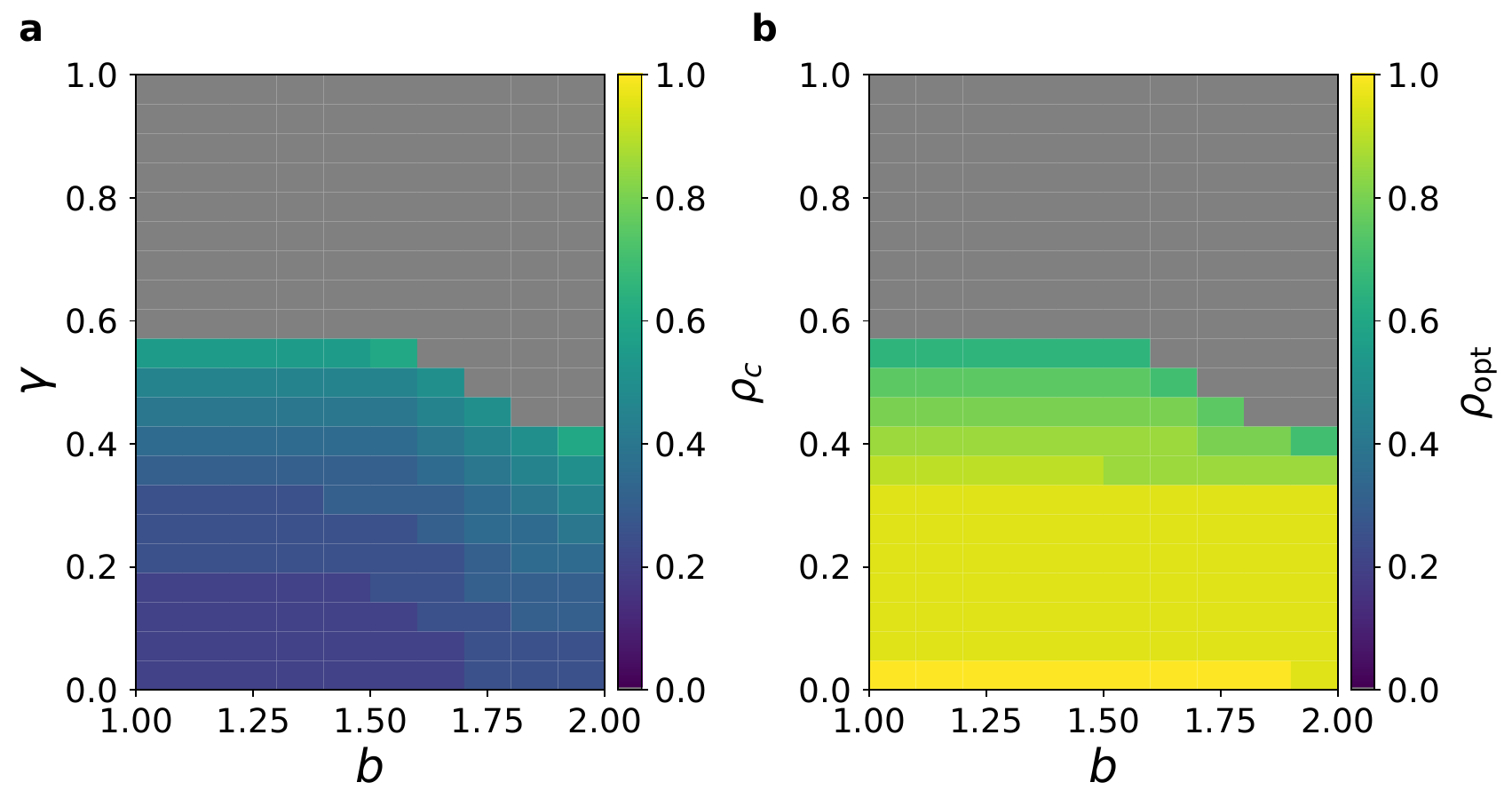}
    \caption{\textbf{Critical and optimal resource investment landscape.} Panel a: Critical resource investment fraction, $\rho_c$ marking the appearance of a finite fraction of fighters in the system, $f_F$, when $\rho>\rho_c$, in control parameter space $(b,\gamma)$. Panel b: Optimal resource investment fraction, $\rho_{\text{opt}}$, marking the maximum fraction of fighters in the system in $(b,\gamma)$-space. Both $\rho_c$ and $\rho_{\text{opt}}$ have been obtained through linear search from the stochastic multi-agent simulations. The gray regions in both plots denote the absence of surviving fighters for any value of $\rho\in[0,1]$.}
    \label{fig:investment_landscape}
\end{figure*}

The previous results show the existence of two interesting values of the invested resource fraction $\rho$, the first of them is the critical value of $\rho$, $\rho_c$, where for $\rho>\rho_c$ and also conditioned on $\gamma$, we observe the coexistence of the cooperative and competitive strategies. The other one is $\rho_{\text{opt}}$, which sets an optimal (maximum) value of the fraction of fighters in the system $f_F$. To obtain a more exhaustive characterization of the system's behavior, we show in Figure \ref{fig:investment_landscape}a the values of $\rho_c$, and in Figure \ref{fig:investment_landscape}b, the values of $\rho_{\text{opt}}$ as obtained from our multi-agent simulations in $(b,\gamma)$-space. That is, for any pair $(b,\gamma)$ in $[1,2]\times [0,1]$, we run simulations scanning for $\rho$ parameter, and by a simple linear search we obtain $\rho_c$ and $\rho_{\text{opt}}$, provided they exist. 

The following observations can be outlined from Figure \ref{fig:investment_landscape}. First, as $\gamma$ increases, $\rho_c$ also increases. Fighters require a smaller investment fraction $\rho$ when the technology parameter $\gamma$ is lower. That is, the more poor-rewarding contests become. Second, further increasing $\gamma$, $\rho_c$ (entering the blank region) both $\rho_c$ and $\rho_{\text{opt}}$ abruptly disappear. Technically, $\rho_c$ is undefined since the phase of fighting-cooperation coexistence does not exist. By the same argument, there is not such an optimal value of $\rho$. Under these conditions, $\gamma>\gamma_c$, for any value of $\rho$ and $b$, fighters are unable to survive in the steady-state no matter the fraction of resources $\rho$ they invest in contest interactions. This leads to a regular PD-scenario, where the system's phase (cooperation, defection, or coexistence of both strategies) will depend on $b$. Third, the dependence of $\rho_{c}$ with varying $b$ is smooth and small, and in fact $\rho_{\text{opt}}$ remains fairly constant across the range of $b$.

In Appendix \ref{app:technology}, for comprehensiveness, the macroscopic behavior of the system is depicted for other values of $\gamma$. Qualitatively, we find the same behavior and phases. In Appendix \ref{app:depletion}, the effect of different resource depletion rates $\alpha$ is analyzed.

The microscopic configurations show that, as $\gamma$ decreases and $\rho$ increases beyond a certain threshold, finite fractions of cooperators and fighters coexist. This reveals an interesting interplay between cooperation and resource-based competition, beyond the coexistence of cooperation and defection in the classical PD beyond $b_c$.

The emergence of fighters in Phase II is related to local neighborhood effects. In the absence of defectors, which are negligible in this regime, configurations are described by the number of cooperators and fighters surrounding each focal agent. This interpretation is supported by qualitative observations of the system's microscopic evolution.

Furthermore, as we have observed in the general macroscopic analysis of Section \ref{subsec:macroscopic}, another interesting nonlinearity arising in the evolutionary dynamics is the existence of an optimal value of $\rho$, $\rho_{\text{opt}}$, that establishes a maximum fraction of fighters in the system. That is, once we are in the cooperation-competition coexistence phase (Phase IV), increasing $\rho$ favors the expansion of the fighter strategy in the system, but a point is reached where further increases of $\rho$ act in detriment, and the fraction of fighters diminishes. The rationale behind this phenomenon is that, while in the beginning, increments of $\rho$ tend to benefit the prosperity of fighters, continuing to increase $\rho$ ends up becoming a very risky situation, because as gains are proportional to the investment fraction, so are losses. If the probability of winning, as given by the Tullock contest function does not scale up in the same way, defeats become detrimental enough to cause a dwindle in the fighters population. 

\section{Discussion and Conclusions}
\label{sec:discussion}

In this work, we have extended the Prisoner's Dilemma Game (PDG) to incorporate genuine resource-based competition, modeled through the Tullock contest success function from Contest Theory. This extension, constituting what we have named as the cooperation-defection-fighting (CDF) model, allows us to explore the interplay between cooperative, exploitative, and purely competitive behaviors in a spatial evolutionary framework.

We have characterized the system’s behavior by analyzing the macroscopic population fractions of each strategy and defining distinct equilibrium phases based on their coexistence. Our results reveal up to five different phases: full cooperation (Phase I), coexistence of cooperation and defection (Phase II), absence of cooperation (Phase III), coexistence of cooperation and competition (Phase IV), and full coexistence of all three strategies (Phase V). For $\rho<\rho_c$, the system behaves similarly to the standard PDG, with the key distinction that for sufficiently high temptation values ($b\gtrsim 1.75$), the depletion of cooperators leads to resource exhaustion, pushing the system into a ``dead economy'' state with no wealth generation. Notably, this phase persists even for $\rho>\rho_c$ when $b\gtrsim 1.75$. Phase IV emerges for $\rho>\rho_c$, where contests become profitable enough for fighters to sustain themselves through cumulative payoffs. This phase extends well beyond the classical critical temptation value $b_c$, which typically marks the point where cooperation begins to decline in a standard PDG. Our results indicate that under certain profitability conditions, cooperation is no longer a monolithic strategy, yet it remains resilient against exploitation. Furthermore, we observe a nontrivial nonlinear effect within Phase IV: increasing investment in contests initially promotes fighter survival, reaching an optimal investment fraction $\rho_{\text{opt}}$, beyond which further investment leads to a decline in the fighter population. This result highlights the risks of excessive resource allocation toward competition.

Our results also show that resource redistribution through contest dynamics can enable the persistence of fighters—resource hoarders that do not contribute to the commons—under specific structural and dynamical conditions. Cooperation remains robust under high redistribution regimes (low $\rho$), but the emergence of competitive subgroups induces new vulnerabilities. These findings open new perspectives on the interaction between cooperation, hierarchy, and strategic exploitation.

Beyond the core analysis, we also investigated how macroscopic outcomes are shaped by variations in the contest technology parameter $\gamma$ and the resource depletion rate $\alpha$. As $\gamma$ increases beyond a critical threshold ($\gamma_c \approx 0.6$), the advantage conferred by resource asymmetry becomes so pronounced that fighters can no longer survive, and the system collapses back into a classical PDG setting dominated by cooperation and defection. At lower values of $\gamma$, contests are less sensitive to wealth disparities, fostering a broader coexistence between cooperators and fighters. Similarly, the depletion parameter $\alpha$ has a non-monotonic effect: both the absence of depletion ($\alpha=0$) and its extreme ($\alpha=1$) lead to the collapse of coexistence phases, either due to unbounded accumulation by competitive agents or the impossibility of sustaining resource advantage. In contrast, moderate depletion supports more balanced dynamics, allowing cooperation and competition to stably coexist. These findings highlight how technological and ecological constraints shape the long-run viability of different behavioral strategies.

Overall, introducing competitive behavior into the spatial evolutionary PDG unveils rich nonlinear dynamics and broadens the spectrum of emergent strategic interactions. While our model remains a simplified representation of real-world behavior, it provides an insightful extension to evolutionary game theory by incorporating explicit competition. Future research could explore more realistic interaction topologies, more sophisticated decision-making processes regarding resource allocation and strategy adoption, or the effects of finite resource availability on these features.

Given the ubiquity of competition in human societies and across the animal kingdom, we believe this extension is not merely a theoretical modeling exercise but a meaningful step toward a deeper understanding of the interplay among the diverse landscape of animal behavior.

\begin{acknowledgements}
The authors thank H. Xia for helpful discussions and comments during the early stages of this work. AdM and YM were partially supported by the Government of Aragon, Spain, and ERDF "A way of making Europe" through grant E36-23R (FENOL). YM acknowledges support from Grant No. PID2023-149409NB-I00 from Ministerio de Ciencia, Innovación y Universidades, Agencia Española de Investigación (MICIU/AEI/10.13039/501100011033) and ERDF ``A way of making Europe''. CGL was partially supported by Departamento de Ciencia, Universidad y Sociedad del Conocimiento, from the Gobierno de Aragón, Spain (Research Group S04$\_$23R). CS was supported by the National Natural Science Foundation of China (Nos:72371052); China Scholarship Council, China under grant (Nos:202306060151). The authors acknowledge the use of the computational resources of COSNET Lab at Institute BIFI, funded by Banco Santander through grant Santander-UZ 2020/0274, and by the Government of Aragon (FONDO–COVID19-UZ-164255). The funders had no role in the study design, data collection, analysis, decision to publish, or preparation of the manuscript.
\end{acknowledgements}

\newpage
\appendix
\renewcommand{\thefigure}{\Alph{section}\arabic{figure}}


\section{Tullock Contest Success Function}
\label{app:tullock}

\setcounter{figure}{0}

Conflicts are not always amenable to reaching an agreement or peaceful solution, and ``win or lose'' scenarios often emerge as the way out to their resolution. A useful, simple probabilistic description of the expected outcome of combat is provided by the formalism of contest success functions (CSF). A CSF \cite{skaperdas1996contest} is a function of the quantified efforts, or resources, invested by the contenders, that gives the probability of winning the contest. Though CSFs are in general defined for a number of contenders larger than two, in the main text as well as in this exposition, we restrict ourselves to pairwise contests. 

Let $r_i$ be the resources of contender $i$, what could be considered as the focal player from the point of view of simulations, and $r_j$ those of contender $j$, one of their neighbors. The CSF function called Tullock, for a positive parameter $\gamma$, meets the requirement that the winning probability $p$ of contender $i$ is invariant under the re-scaling of both contenders' resources, i.e., for all $\lambda > 0$, $p(\lambda r_i, \lambda r_j)=p(r_i,r_j)$. Explicitly, the Tullock function:

\begin{equation}
\label{eq:TullockFunction}
P_{CSF}(r_i,r_j;\gamma)=\frac{r_i^\gamma}{r_i^\gamma+r_j^\gamma},
\end{equation}
gives the winning probability of contender $i$. A basic assumption behind this result is that victory and defeat (from a contender perspective) are a mutually exclusive complete set of events so that $p_{\gamma}(r_i, r_j) = 1-p_{\gamma}(r_j,r_i)$. 

\begin{figure}[h]
\centering
    \includegraphics[width=0.5\textwidth]{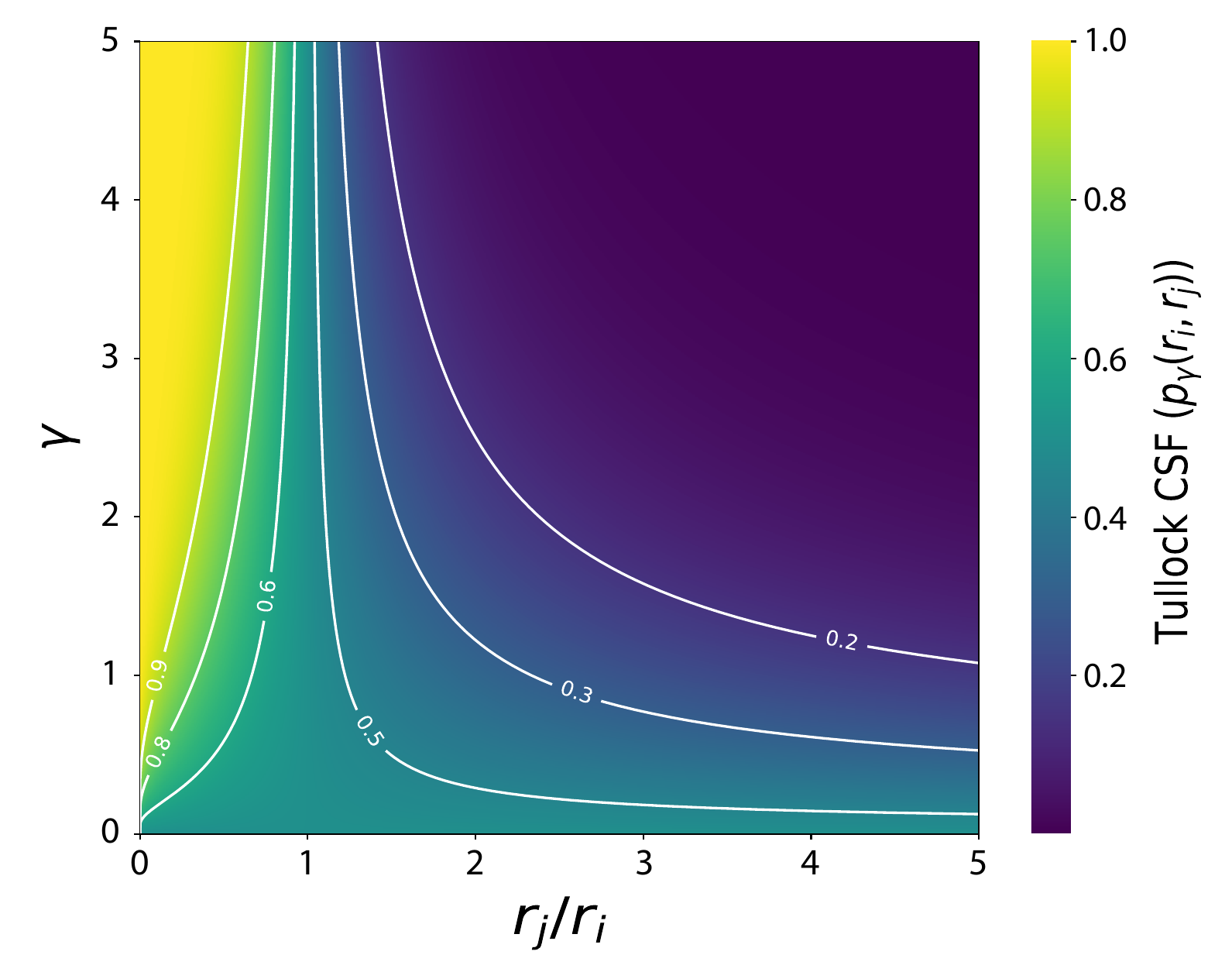}
    \caption{
    \textbf{Tullock Contest Success Function.} Heatmap for the Tullock CSF, $p_{\gamma}(r_i,r_j)$ as a function of the contenders' resource ratio $r_j/r_i$ and the technology parameter $\gamma$. The Tullock CSF yields the probability of player $i$ winning the contest over player $j$ based on the resources deployed by every contender and weighted by the technology parameter $\gamma$.}
    \label{fig:fig_app_heatmap_tullock}
\end{figure}

Regarding the consequences of the contest outcome, one assumes that the winner's benefits are the sum $r_i+r_j$ of both resources, and the loser obtains nothing, zero benefits. The parameter $\gamma$ of the Tullock CSF turns out to play a very important role, because when $\gamma>1$, it is easy to see that whenever $r_i>r_j$, the expected gain for contender $i$ after the combat, $p_{\gamma}(r_i,r_j) (r_i+r_j)>r_i$, and then the (richer) contender $i$ has an incentive to fight, while if $\gamma<1$, the expected gain for the richer contender is lower than their resources before the combat, $p_{\gamma}(r_i,r_j) (r_i+r_j)<r_i$, and thus it is the poorer contender who should rationally decide to fight. 

Following the acutely descriptive terms introduced in \cite{dziubinski2021strategy}, we will call {\it rich-rewarding} a Tullock CSF with parameter $\gamma>1$, and {\it poor-rewarding} a Tullock CSF with $\gamma<1$. In this reference, \cite{dziubinski2021strategy}, where contests refer to events of ``real'' war among nations, a conventional war would be described by a rich-rewarding CSF, while guerrilla warfare would better be described by a poor-rewarding CSF Tullock function, which led the authors to refer to $\gamma$ as ``technology parameter'', and ponder its relevance to the expectations and chances for peaceful coexistence among nations or coalitions. Correspondingly, in economic contests, a rich-rewarding CSF corresponds to a competition in a conventional costly scenario and a poor-rewarding CSF to either a low-cost strategy or a guerrilla marketing scenario.

\section{Macroscopic behavior for varied technology parameter}
\label{app:technology}

\setcounter{figure}{0}

\begin{figure*}
\centering
    \includegraphics[width=1.0\textwidth]{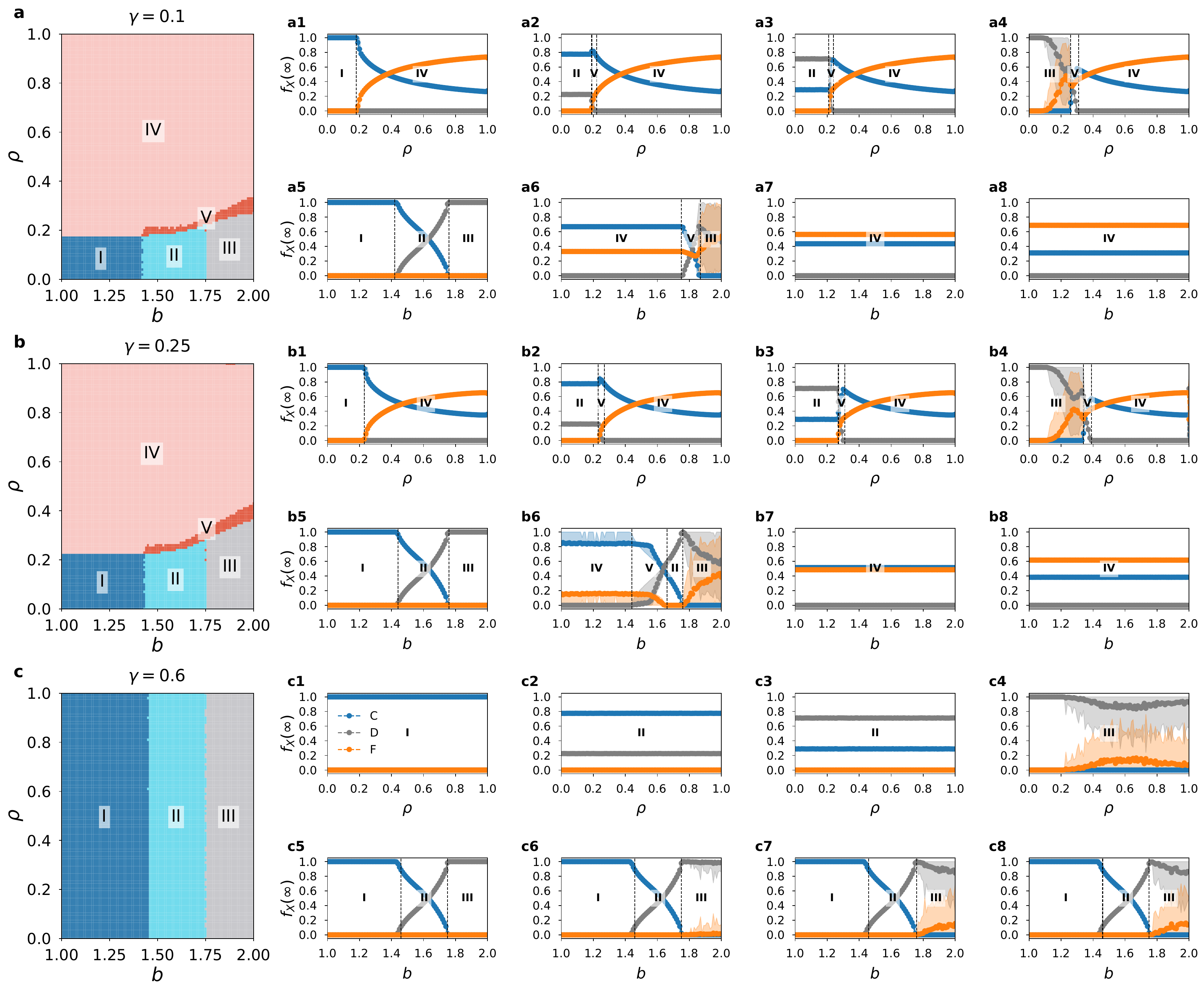}
    \caption{\textbf{The effect of different poor-rewarding technology parameters.} Panels a, b, and c depict the system macroscopic phases on $(b,\rho)$-space, for technology parameter $\gamma=0.1$, $0.25$, and $0.45$, respectively. Phases are defined in the following way. Phase I: Full Cooperation, $\langle f_D(\infty)\rangle=\langle f_F(\infty)\rangle=0$. Phase II: Cooperation and defection. Phase III: Dead economy, defined by $\langle f_C(\infty)\rangle=0$, and thus $\Pi(\infty)_i\to 0$ for any other agent $i$ surviving in the long run. Phase IV: Cooperation and competition, $\langle f_F(\infty)\rangle=0$. Phase V: Triple coexistence, where all populations are able to survive at the equilibrium. Accompanying plots represent the strategy fractions. For all $\gamma$ values, panels a1 to a4, b1 to b4, and c1 to c4, correspond to fixed $b$ sections of value $b=1.1$, $1.5$, $1.7$ and $1.9$, respectively. On the other hand, panels a5 to a8, b5 to b8, and c5 to c8, correspond to fixed $\rho$ sections of value $\rho=0.05$, $0.25$, $0.45$, and $0.75$.  Dashed lines represent the different phase transitions that we encounter. Every point in the diagram and curves is the result of averaging over $100$ simulations with a system size of $N= 10^4$, with depletion rate always at $\alpha=0.1$.}
    \label{fig:app_phases_poorgamma}
\end{figure*}

In Section \ref{subsec:nonlinear}, we described the investment landscape ($\rho_c$ and $\rho_{\text{opt}}$) for varying $\gamma$. This already showed us where to expect coexistence between cooperation and competition (and thus Phase IV) for a given $\gamma$, and also whether there was non-monotonic behavior for $f_F(\rho)$ in $\rho\in(\rho_c,1)$. In this Appendix, we complement the phase diagram shown in Figure \ref{fig:phases_master} by exploring the system's macroscopic behavior for other values of the technology parameter $\gamma$ (Figure \ref{fig:app_phases_poorgamma}).

Accordingly, we showcase the macroscopic behavior in $(b,\rho)$-space for $\gamma=0.1$ (Panel a), $0.25$ (Panel b) and $0.6$ (Panel c). Complementarily, the strategy fractions as functions of $\rho$ for fixed $b$, and conversely, are also represented. For all $\gamma$ values, panels a1 to a4, b1 to b4, and c1 to c4, correspond to fixed $b$ sections of value $b=1.1$, $1.5$, $1.7$ and $1.9$, respectively. On the other hand, panels a5 to a8, b5 to b8, and c5 to c8, correspond to fixed $\rho$ sections of value $\rho=0.05$, $0.25$, $0.45$, and $0.75$.

From the phase diagrams we can highlight the following trends:
\begin{itemize}
    \item By decreasing $\gamma$ from the representative $\gamma=0.45$ showcased in the main text, we observe how Phase IV expands towards lower $\rho$, that is, $\rho_c$ decreases. In turn, this expansion makes both Phase V - Triple Coexistence- and Phase III - Dead Economy- to shrink and be restricted to a much smaller region in $(b,\rho)$ control parameter space. Regarding $\rho_{\text{opt}}$, we also observe how it has been pushed towards $\rho_{\text{opt}}\to 1$, and thus no non-monotonous behavior is observed for lower $\gamma$.
    \item By increasing $\gamma$ from $0.45$ to $0.6$, we effectively observe the disappearance of Phase IV - cooperation and competition, as well as of Phase V, as expected from what we obtained in Figure \ref{fig:investment_landscape}. 
\end{itemize}

Even though we have previously seen how when $\gamma>\gamma_c\approx 0.6$, fighters cannot survive in the equilibrium and cooperation becomes hegemonic as the dominant strategy, for completeness, we extend the exploration of the role of the technology parameter $\gamma$ to the so-called rich-rewarding regime, $\gamma>1$ (see Appendix \ref{app:tullock}).

\begin{figure}
\centering
    \includegraphics[width=0.5\textwidth]{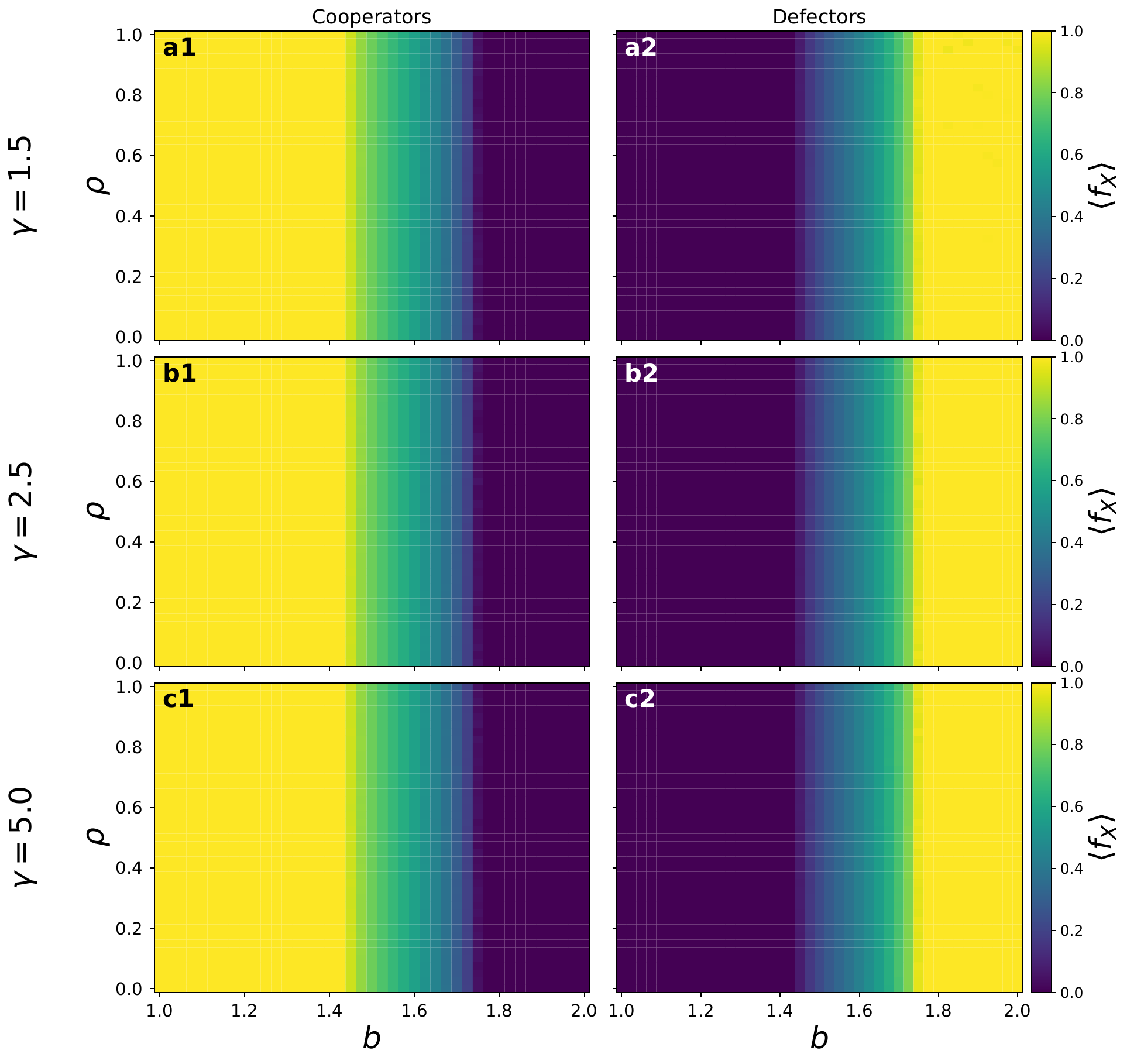}
    \caption{
    \textbf{System behavior in the rich-rewarding technology regime}. Steady-state average fraction of cooperators (panels a1, b1, c1) and defectors (panels a2, b2, c2) across values of the technology parameter $\gamma$ in the rich-rewarding regime: $\gamma = 1.5$ (a1, a2), $\gamma = 2.5$ (b1, b2), and $\gamma = 5.0$ (c1, c2). The fighting strategy is omitted as it fails to persist at steady state for any $(b, \rho)$ and any $\gamma > 1$. As a result, classical weak PDG solutions are recovered along $b$, independently of $\rho$ and $\gamma$.}
    \label{fig:app_heatmap_richgamma}
\end{figure}

Figure \ref{fig:app_heatmap_richgamma} represents now directly a collection of heatmaps for the strategy population fractions at equilibrium in $(b,\rho)$-space, for technology parameters $\gamma=1.5$ (panels a1 and a2), $\gamma=2.5$ (b1 and b2), and $\gamma=5.0$ (c1 and c2). In all cases, the fighter strategy fails to survive at equilibrium and is therefore omitted for clarity. Now, the system exhibits three distinct phases, Phase I (Full Cooperation), Phase II (Cooperation and Defection), and Phase III (Dead Economy), each extending vertically across $\rho$, i.e., independent of the value of $\rho$.

As shown in Appendix \ref{app:tullock}, the Tullock Contest Success Function operates in such a way that tends to promote the victory of the contender with more resources. This effect is exacerbated as $\gamma$ increases. However, we have found through our work that cooperators, on average, tend to be wealthier than fighters; thus, when facing a contest from a neighboring fighter, the probability of the poorer fighter winning significantly decreases as $\gamma$ increases. 

Thus, in the rich-rewarding regime, we find a rather classical weak PDG scenario (with cumulative resources) where the third strategy, fighters, and the associated parameters of the contest interaction, $\rho$ and $\gamma$, do not play any role. For $b<b_c$, full cooperation reigns. For $b>b_c$, defectors start to populate the system at the expense of cooperators. When the temptation parameter grows high enough, cooperation disappears, and the system is dominated by defectors. It must be noted that, as $\alpha\neq 0$ in this exploration and cooperation has disappeared, the full defection phase is equivalent here to the 'dead economy' phase described in the main text. 

\section{The effect of resource depletion}
\label{app:depletion}

\setcounter{figure}{0}

In our model, agents accumulate payoffs over time, which can grow or decline depending on interactions. This accumulation is moderated by the resource depletion parameter $\alpha$, which effectively discounts past earnings at each round. Note that payoffs can be negative depending on contest outcomes, so accumulation is not strictly increasing.

\begin{figure*}
\centering
    \includegraphics[width=1.0\textwidth]{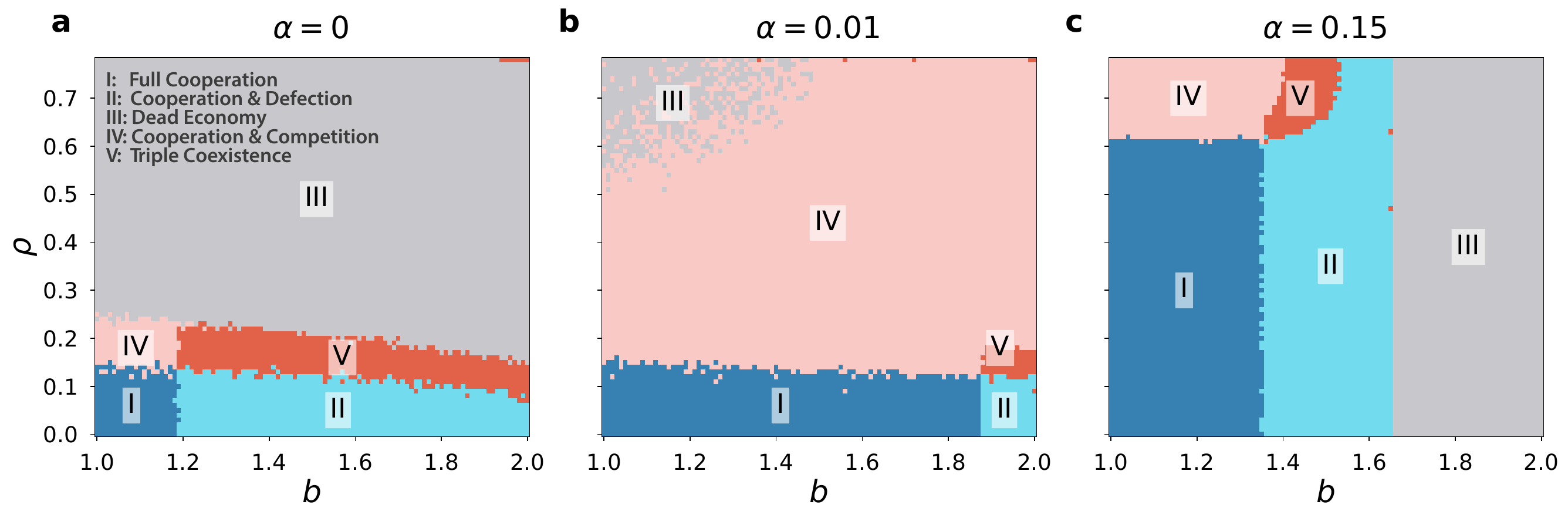}
    \caption{
    \textbf{The effect of resource depletion rates.} Phase diagrams in the $(b,\rho)$ space for varying values of $\alpha$: $\alpha=0$ (a), $\alpha=0.01$ (b), and $\alpha=0.15$ (c). All simulations use $\gamma=0.45$ and a system size of $N=10^4$, averaged over $100$ runs per parameter setting.}
    \label{fig:app_phase_depletion}
\end{figure*}

In the main text, we focused on a moderate depletion rate of $\alpha=0.1$, which gives some weight to historical payoff accumulation while still allowing current performance to matter. Here, we extend the analysis by exploring how varying $\alpha$ affects macroscopic behavior. Specifically, we present phase diagrams in the $(b, \rho)$ control parameter space for three different depletion rates: $\alpha = 0$, $0.01$, and $0.15$ (see Fig.~\ref{fig:app_phase_depletion}). In all cases, the Tullock contest technology parameter is fixed at $\gamma=0.45$.

When there is no resource depletion ($\alpha=0$, Fig.~\ref{fig:app_phase_depletion}a), all five macroscopic phases identified in the main text for $\alpha=0.1$ remain present, but their domains shift significantly. Notably, the 'Dead Economy' phase (Phase III) expands to dominate the upper region of the diagram ($\rho > 0.25$) across the full range $b \in [1, 2]$, becoming the prevalent solution. The triple coexistence phase (Phase V) now appears as a narrow horizontal band, flanked by Phase III above, Phase II below, and Phase IV to the left. This region of mixed cooperation and competition is drastically reduced compared to its extent at $\alpha=0.1$, as is the full cooperation phase (Phase I), which is confined to a small corner. In contrast, Phase II has expanded horizontally and now covers a broader range of $b$ values, up to $b=2$.

It is worth highlighting that in the absence of depletion, Phase III is populated exclusively by fighters (rather than by a mix of fighters and defectors). In a sense, the lack of resource decay strongly favors the accumulation-driven success of competitive strategies. However, this comes at a systemic cost: cooperation collapses, and with it the economy itself, as the only wealth generators vanish. Thus, although fighters thrive when $\alpha=0$, they do so at the expense of long-term system viability.

Introducing a small but nonzero depletion rate ($\alpha=0.01$, Fig.~\ref{fig:app_phase_depletion}b) alters the landscape substantially. The triple coexistence phase (V) is pushed into a small corner with $\rho \in (0.1, 0.2)$ and $b \in (1.9, 2)$. However, the other solutions involving cooperators expand considerably. Phase I extends up to $b \approx 1.9$ for $\rho < 0.2$, well beyond the classical Prisoner's Dilemma regime and the reference case of $\alpha=0.1$. This indicates a non-monotonic relationship between $\alpha$ and cooperative success: cooperation is actually promoted by moderate depletion. Furthermore, Phase IV becomes dominant, relegating Phases III and V to marginal roles. This shows that not only cooperation but also competitive strategies may optimally thrive at intermediate values of $\alpha$.

Finally, increasing depletion slightly above the main text reference to $\alpha=0.15$ (Fig.~\ref{fig:app_phase_depletion}c) leads to a contraction of coexistence phases. Phases I and II expand vertically to higher $\rho$ values, while Phases IV and V become confined to narrow high-$\rho$ domains. This trend suggests that larger $\alpha$ progressively eliminates the mixed-strategy equilibria involving fighters, favoring the recovery of the classical PD landscape.

In the extreme case $\alpha=1$ (not shown here), trivially, all accumulated payoffs are entirely erased at each round, reducing the game to a series of one-shot interactions with no effective memory. In this regime, competitive (fighter) strategies are untenable, as they would always enter contests with zero resources, rendering them unviable. In contrast, when $\alpha=0$, agents retain the full history of their accumulated payoffs, effectively summing the results of all past rounds. This corresponds to a model where memory is perfect ($M \to \infty$), and the strategy update decisions reflect the entire trajectory of past performance. 

In the extreme case $\alpha=1$ (not shown), all accumulated payoffs are entirely erased at each round, reducing the game to a series of one-shot interactions with no effective memory. In this regime, competitive (fighter) strategies are untenable, as they always enter contests with zero resources. By contrast, the $\alpha=0$ limit represents perfect memory ($M \to \infty$), where agents retain the full history of past payoffs. Taken together, these extremes illustrate how resource depletion acts as a temporal filter: too little encourages exploitative dominance; too much erases the benefits of past investment. Between these poles, however, lies a narrow window where both cooperation and healthy competition can coexist.

\bibliography{references}

\end{document}